\documentclass{article}
\usepackage[T1]{fontenc} % add special characters (e.g., umlaute)
\usepackage[utf8]{inputenc} % set utf-8 as default input encoding
\usepackage{ismir,amsmath,cite,url}

\usepackage{graphicx}
\usepackage{color}
\usepackage{tabularx}
\usepackage{subcaption}
\usepackage{soul}
\usepackage{boldline}
\usepackage{multirow}

\usepackage{cite}
\usepackage{booktabs}

\newcommand{\jordan}[1]{\textcolor{black}{#1}}
\newcommand{\paul}[1]{\textcolor{black}{#1}}

\newcommand{\yh}[1]{\textcolor{black}{#1}}

\title{Neural Loop Combiner: Neural Network Models for Assessing the Compatibility of Loops}

\multauthor{Bo-Yu Chen$^{1}$,~ Jordan B. L. Smith$^2$, and Yi-Hsuan Yang$^{1}$} {  
  $^1$ Academia Sinica, Taiwan,~
  $^2$ TikTok, London, UK\\
{\tt\small bernie40916@citi.sinica.edu.tw,  jordan.smith@tiktok.com,  yang@citi.sinica.edu.tw }
}
\def\authorname{Bo-Yu Chen, Jordan B. L. Smith, Yi-Hsuan Yang}

\begin{document}

\maketitle
\begin{abstract}
% Music producers often base their work on loops and they may have access to libraries of thousands. However, finding loops that are compatible is a time-consuming process;
\jordan{Music producers who use loops may have access to thousands in loop libraries, but finding ones that are compatible is a time-consuming process;}
we hope to reduce this burden with automation. State-of-the-art systems for estimating compatibility, such as AutoMashUpper, are mostly rule-based and
\jordan{could be improved on with}
% do not benefit from the power of modern
machine learning. % 
\jordan{To train a model, we need a large set of loops with ground truth compatibility values. No such dataset exists, so we extract loops from existing music to obtain positive examples of compatible loops, and propose and compare various strategies for choosing negative examples.}
% In the absence of a dataset of ground truth compatibility values, we propose to obtain the required data by extracting loops from existing music in order to obtain positive examples of compatible loops, and we propose and compare several strategies for collecting negative examples.
For reproducibility, we curate data from the Free Music Archive. Using this data, we investigate two types of model architectures for estimating the compatibility of loops: one based on a Siamese network, and the other a pure convolutional neural network (CNN). We conducted a user study in which participants rated the quality of the combinations suggested by each model, and found the CNN to outperform the Siamese network. Both model-based approaches outperformed the rule-based one. We have opened source the code for building the models and the dataset.
% , to facilitate follow-up research on loop analysis and creation.
\end{abstract}

\section{Introduction}
\label{sec:introduction}

The emergence of digital audio editing techniques and software such as digital audio workstations (DAWs) has changed the way people make music. In a DAW, producers can easily produce music by making use of pre-existing audio as loops. Loops are used in many styles, especially Electronic and Hip-Hop music.
% because of their seamless and repeatable characteristic.
% The causality of this seems suspect --- are they used BECAUSE they are seamless, or are they made to be seamless because they are going to be used as loops?
Perhaps noticing the business value of such loop-based music, many companies such as Splice and LANDR have built up databases of loops.
Community efforts have flourished too,
including Looperman, an audio community that provides free loops for  music producers to \jordan{use}.
%\st{create music out of.}
% Therefore, the producer can comfortably obtain sufficient resources from the Internet to express their thought.
But having so many resources to choose from leaves a separate problem: how to navigate the library efficiently and how to choose which loops to combine.
These %\st{are tasks requiring}
\jordan{tasks require} human practise, expertise, and patience:
% However, it is still difficult still not easy for the novice producer to find suitable loop ingredients from the large-scale loops database.
%the experienced producer
%\st{has the ability to}
to 
%\jordan{can} 
recognize the characteristics of the loop and to determine whether it is %\st{usable}
\jordan{suitable} for their song.
% The only way to cultivate this music intuition is through long-term accumulation.
However, thanks to recent advances in music information retrieval (MIR), we believe it is possible to make loop selection from large-scale loops database easier.

A few existing systems could potentially solve this problem.
%\st{For example, }
Kitahara \emph{et al.}~\cite{15smc_sequnecer} presented an intelligent loop sequencer that \jordan{chooses loops to fit a user-defined `excitement' curve,}
% \st{can select the loop automatically, and proposed to choose loops based on their level of `excitement.'}
but, \jordan{excitement}
%\st{probably}
only accounts for part of what makes two loops compatible.
% \paul{One of the previous study \cite{15smc_sequnecer} tackle the same problem with us, while the study [cite sequencer] propose to choose the loops based on their excitement which may be the part of compatibility but not whole. Another study\cite{14aslp_atp}, the study proposes an interactive mashup creation system for combining multiple songs.}
The AutoMashUpper system~\cite{14aslp_atp} could also be applied to the loop selection process: it involves estimating the \jordan{`mashability' of two songs (i.e., how compatible they would be if played at the same time)}. %\st{mashability of two songs in order to decide whether to combine them}. 
It is a rule-based system that computes the harmonic and rhythmic similarity and spectral balance, and it has proven useful in other applications \cite{15smc_music_puzzle,19ismir_voice}. 
However, AutoMashUpper
% \cite{14aslp_atp}
% is successful for the previous study, there are
has two limitations that could be improved by current machine learning methods. First, it regards the harmonic and rhythmic similarity as part of mashability, while it is actually possible that two music segments match perfectly despite having different harmonies and rhythms. Second, hand-crafted representations cannot fully describe all features in the music segment. 

To capture the more complicated compatible relationship between two music segments, we propose to employ %\st{the power of} 
modern machine learning models to learn to predict the compatibility of  loops. 
%\st{The main obstacle that prevents the development of such a model, according to our knowledge, is mainly}
\jordan{A major obstacle preventing the development of such a model is} the lack of a dataset with sufficient %\st{number of} 
labelled data. While there are \jordan{many} datasets of loops, none %\st{of them} 
provide %\st{the} 
ground truth compatibility values. 

% The major contribution of this paper is two-fold.
\jordan{We make two main contributions.}
%\begin{itemize}
%    \item 
    %\st{In consequence,} 
    First, %\st{primary} 
    we propose a data generation pipeline that supplies labelled data for model-based compatibility estimation (see Section 3). This is done by using an existing loop extraction algorithm %proposed %\st{fairly} 
    %recently 
    \cite{18icassp_loop_sepearation} to yield positive examples of loops that have been used together in real loop-based music. 
    % a lot of loop pairs used to combined before by recently loop extraction algorithm and directly learn the compatibility of two loops from these loop pairs.}
    We also propose %loop refinement and loop pairing
    procedures to ensure the quality of the positive data, and investigate different strategies to mine negative data from the result of loop separation. 

%    \item 
    Second, we develop and evaluate two neural network based models for loop compatibility estimation (see Section 4),
    %, using data from the aforementioned data generation pipeline. 
    %Specifically, we consider two %\st{fairly} 
    %generic and well known architectures for this work,
    one based on convolutional neural network (CNN) and the other Siamese neural network (SNN) \cite{94siamese}. The two approaches perform loop compatibility estimation using %\st{fairly} 
    different approaches: the CNN directly evaluates the combination of loops in a time-frequency representation, whereas SNN processes the two loops to be evaluated for compatibility separately.  We report both  objective and subjective evaluations (see Section 5) to study the performance of these approaches, and to %validate the superiority 
    compare the model-based approaches \jordan{with} AutoMashUpper.

%\end{itemize}

The audio data we employ to build the neural network is from the Free Music Archive (FMA) \cite{17ismir_fmadataset} (see Section \ref{sec:date_generation:data}), which make data re-distribution easier. Moreover, we %\st{also plan to open source}
\jordan{have open-sourced} the code implementing the proposed CNN and SNN models at \url{https://github.com/mir-aidj/neural-loop-combiner}.

% \paul{In sum, the major contribution in this paper is the methodology to create positive and negative data for training supervised neural networks. The overall methodology contains the data generation pipeline which can effectively create positive training data, negative sampling strategies which capable to make reasonable incompatible negative data, and the two proposed models which make use of the training data created by the above methods.}

\section{Related work}
\label{sec:related_work}
%\st{Because of} 
\jordan{Along with} the growing interest in loop-based music, academic studies focusing on assisting loop-based music production have become popular. An early study \cite{08icme_loop_explore} proposed two ways to help create loop-based music:
automatic loop extraction and assisted loop selection. %\st{which build up} 
\jordan{This laid} the foundation for this field.

For \textbf{loop extraction}, Shi and Mysore \cite{18chiloopmaker} proposed an interface with automatic and semi-automatic modes for the producer to find the most suitable segment in a piece of music to excerpt and use as a loop, by cropping directly.
These algorithms
%for %\st{selecting}
%\jordan{extracting} segments 
\jordan{estimate}
% are based on rule-based
similarity with handcrafted features: harmony, timbre, and energy~\cite{08icme_loop_explore, 18chiloopmaker}. However, \jordan{the segments are excerpted without any attempt to isolate one part of a potentially multi-part piece of music.}
\jordan{One solution to this used a common quality of loop-based music---that loops are often introduced one at a time---to recover the source tracks~\cite{seetharaman2016}, but not all pieces have this form.}
%\st{loop-based music usually has multiple layers of music, making the performance of such a cropping approach limited.}
To tackle both problems, Smith \emph{et al.} \cite{18icassp_loop_sepearation,19ismir_unmixer} proposed to extract loops by taking  into account how they repeat. The algorithm they proposed can directly extract the loops from songs using nonnegative tensor factorization (NTF).

%These systems for improving the loop extraction step are crucial, but loop extraction is only half of the problem: 
%To form meaningful loop-based music, \textbf{loop selection} is also needed.
% However, to form meaningful loop-based music, the loop selection process is also crucial.  There have been a number of recent studies focusing on improving the performance of loop extraction, yet few of them tackle the loop selection problem.
For \textbf{loop selection}, 
Kitahara \emph{et al.} \cite{15smc_sequnecer} proposed to select the loops according to the level of excitement entered by the user. However, we see three limitations in this work. First, the study focuses on Techno music only---while excitement is certainly highly relevant to this genre, the approach may not generalize well to other genres.  Second, the level of excitement has to be manually entered by a user, which limits its usability. 
%In other words, to find a proper loop, a user needs to clarify what degree of excitement should be entered. 
Third, the study does not take compatibility into consideration. As a result, even though a user can find loops with the desired excitement level, the loops may not be compatible with one another.

To the best of our knowledge, the work of Davies \emph{et al.}~\cite{13ismir_atp,14aslp_atp} represents the first study to investigate mashability estimation.
%\st{or how to automatically find musical fragments that could be combined with a query song. Specifically, }
%to compute the mashability between two music segments, 
Their %\st{resulting} 
AutoMashUpper system represents each music segment (not necessarily a loop) with a chromagram, a rhythmic representation, and a spectral band representation, each made to be beat-synchronous.
% as a beat-synchronous chromagram, a beat-synchronous rhythmic representation, and a beat-synchronous spectral representation.
Given two songs, it computes the similarity between the chromagrams and the rhythmic representations, and the spectral balance between the songs, to obtain the final mashability estimate.
% Although it is a rule-based approach, it is general-purpose and represents 
%the most relevant work for our system and
\yh{While AutoMashUpper is developed for general mashability estimation of longer music pieces, it can also be applied to loop selection.}
%Moreover, for the purpose of loop section, it may be more general than the algorithm proposed by Kitahara \emph{et al.} \cite{15smc_sequnecer}.

%\paul{Besides AutoMashUpper, other studies extend their algorithms. However, none of them are as general as AutoMashUpper.} 
% \jordan{Extensions of AutoMashUpper include the work of Lee \emph{et al.} \cite{15ismir_mashup_vh}, who proposed that, besides having similar harmony, two parts should be compatible if they have complimentary amounts of harmonic instability (i.e., a clip with varying harmony should be paired with a clip with steadier harmony).
% In a series of works, Bernardes \emph{et al.}\cite{15DAFx_Mixing,16AppledSciences_Mixing,17CMMR_Mixing,17CMMR_H_Mixing}  
% proposed several harmonic mixing approaches based on psychoacoustic principles and ``tonal interval space indicators.'' 
% Xing \emph{et al.} \cite{20_popmash} used paired audio, MIDI, and lyrics data for mashability estimation, taking into account melodic, rhythm, and lyrical rhyme similarity. 
% \yh{Although these alternatives have useful aspects, AutoMashUpper remains a good general-purpose approach, with code publical available, and requires only audio data. We hence consider it as the main baseline.}
% }

\jordan{Lee \emph{et al.} \cite{15ismir_mashup_vh} extended AutoMashUpper, proposing that two parts should be more compatible if they have complimentary amounts of harmonic instability.
% (i.e., a clip with varying harmony should be paired with a clip with steadier harmony).
They generated mashups that were preferred by listeners to the output of AutoMashUpper,
but they focused on a particular type of mash-up (combinations of vocal and backing tracks) whereas we focus on general loop compatibility.}
% Extensions of AutoMashUpper has been made.
% %though none of them is as general as AutoMashUpper.
% Lee \emph{et al.} \cite{15ismir_mashup_vh} studied the ``vertical'' aspect of mashabiltiy estimation, i.e., finding suitable lead units to be overlaid with background units. %This work is not relevant to loop selection.
% As a result, we consider AutoMashUpper as the baseline model in this paper.
%\paul{Bernardes \emph{et al.}\cite{15DAFx_Mixing} \cite{16AppledSciences_Mixing} \cite{17CMMR_Mixing} \cite{17CMMR_H_Mixing} 
%In a series of works, 
Bernardes \emph{et al.}\cite{16AppledSciences_Mixing,17CMMR_Mixing}  
proposed several harmonic mixing approaches based on psychoacoustic principles and ``tonal interval space indicators.'' 
%However, these approaches focused on harmonic compatibility rather than general mashability.
%Harrison \jordan{and Pearce} \cite{20Psy_consonance} developed a computational model that to predict musical chord’s simultaneous consonance. 
%Nevertheless, the computational model is modeling from Western music which may not suitable for loop-based music.
Xing \emph{et al.} \cite{20_popmash} used paired audio, MIDI, and lyrics data for mashability estimation, taking into account similarity in melody, rhythm, and rhyming lyrics. 
% to select the suitable music fragment. However, to conduct the music similarity algorithms, The pair MIDI, waveform, and lyrics data is necessary. 
%To our best knowledge, there are not any public loop-based music dataset contains three types of data in the meantime.  
%Therefore, although the AutoMashUpper is not a specific design for estimating the mashability of loops, It is general-purpose and represents. Furthermore, if it works for long chucks, there might also work well for short chucks. Thus, we believe there are no other algorithms that are better suited for AutoMashUpper as the baseline model in this paper.}
\jordan{Among these works, AutoMashUpper stands out as a general-purpose system requiring only audio data, so we consider it as our baseline.}
% \yh{Although these extensions are interesting, AutoMashUpper stands out as it is general-purpose and requires only audio data. We hence consider it as the main baseline.}

\begin{table}
  \begin{center}
    \begin{tabular}{l|rrr}
    \toprule
      \textbf{Data type} & \textbf{\# loops} & \textbf{\# loop pairs} & \textbf{\# songs} \\
      \midrule
      Training set      & 9,048  & 12,774 & 2,702\\
      Validation set   & 2,355  & 3,195  & 7,06 \\
      Test set       & 200    & 100    & 100  \\
      \midrule
      $\sum$          & 11,603 & 16,069 & 3,508\\
      \bottomrule
    \end{tabular}
    \caption{Statistics of the dataset, which is derived from Hip-Hop songs in the Free Music Archive (FMA) \cite{17ismir_fmadataset} using the data generation pipeline shown in Figure \ref{fig:pipline}.}
    \label{tab:numbers}
  \end{center}
\end{table}

\section{Data Generation Pipeline}
\label{sec:date_generation}

% In this section, we describe the data generation pipeline that can generate valid loop pairs for training a machine learning model. We first introduce the dataset used for our pipeline, then we review the loop extraction algorithm (i.e., \cite{18icassp_loop_sepearation}) we employ to extract loops and layouts from existing songs. After that, we introduce the purification step to retrieve the valid loops and activation of the loops throughout the song. The latter is also referred to as the loop layout. Finally, we retrieve the valid loop pairs by paring valid loops from the valid layouts.
To create the labeled data needed for training 
%\st{machine learning}
\jordan{our} models, we obtained a dataset of loop-based music and used loop extraction \jordan{\cite{18icassp_loop_sepearation}} to obtain audio loops. We use a \jordan{new} loop refinement procedure to reduce the number of duplicate loops per song, and a loop pairing procedure to get \jordan{pairs of loops that co-occur in songs}. 
%\st{loops co-occurred in real music}.

\subsection{Dataset}
\label{sec:date_generation:data}

To extract a collection of loops, we first need a large set of songs that use loops.
We chose to use music from the Free Music Archive (FMA)~\cite{17ismir_fmadataset} \jordan{so that we could redistribute the audio data. We}
% and we 
restricted ourselves to the genre of Hip-Hop \jordan{for three reasons:}
% This \jordan{served} several \st{purposes}\jordan{needs}:
First, we could not manually verify whether each song used loops, so we \jordan{needed} %\st{wanted} 
to use a genre known for using loops.
Second, for the loop extraction step to work, we needed the music to have a steady tempo.
\paul{Third, we expected that a Hip-Hop subset would provide a \jordan{useful} %\st{wide}
variety of loops since the \jordan{genre} %\st{style} 
is known for incorporating loops from many \jordan{other genres}}
%\st{music styles.}
% \st{Finally, it is easier to redistribute the audio data from FMA. }

We collected 6,868 Hip-Hop songs in total from FMA by searching for the tag ``Hip-Hop.'' We passed these songs through the proposed data generation pipeline, and kept the 3,508 songs from which we could find at least %\st{a}
\jordan{one} valid loop pair. Specifically, from these 3,508 songs, we %\st{got}
\jordan{obtained} 11,603 valid loops and 16,069 valid loop pairs.
\jordan{From the full set of loops, we reserved 200 loops (1 pair each from 100 different songs) for the evaluations (see Section \ref{sec:exp and eval}), and then split the rest into train and validation sets in a 4-to-1 ratio (see Table \ref{tab:numbers}).}\footnote{To facilitate follow-up research, we plan to share publicly the loops we extracted from FMA songs. While FMA has looser copyright restrictions, the songs are associated with 7 different Creative Commons licenses with various degree of freedom in data manipulation and re-distribution. We will therefore build up our dataset into 7 groups, one for each unique license, before distributing the  loops.}
%\st{We then split the data into training, validation and test sets, as shown in Table \ref{tab:numbers}. 
%\paul{To evaluate the model, we first picked up the 100 loop pairs, which are entirely independent of the other sets. Then, Reminders split into training and validation sets with a 4-to-1 ratio.}}

\begin{figure}[t]
\centering
%\vspace{-1em}
\includegraphics[width=7cm,height=7cm,keepaspectratio]{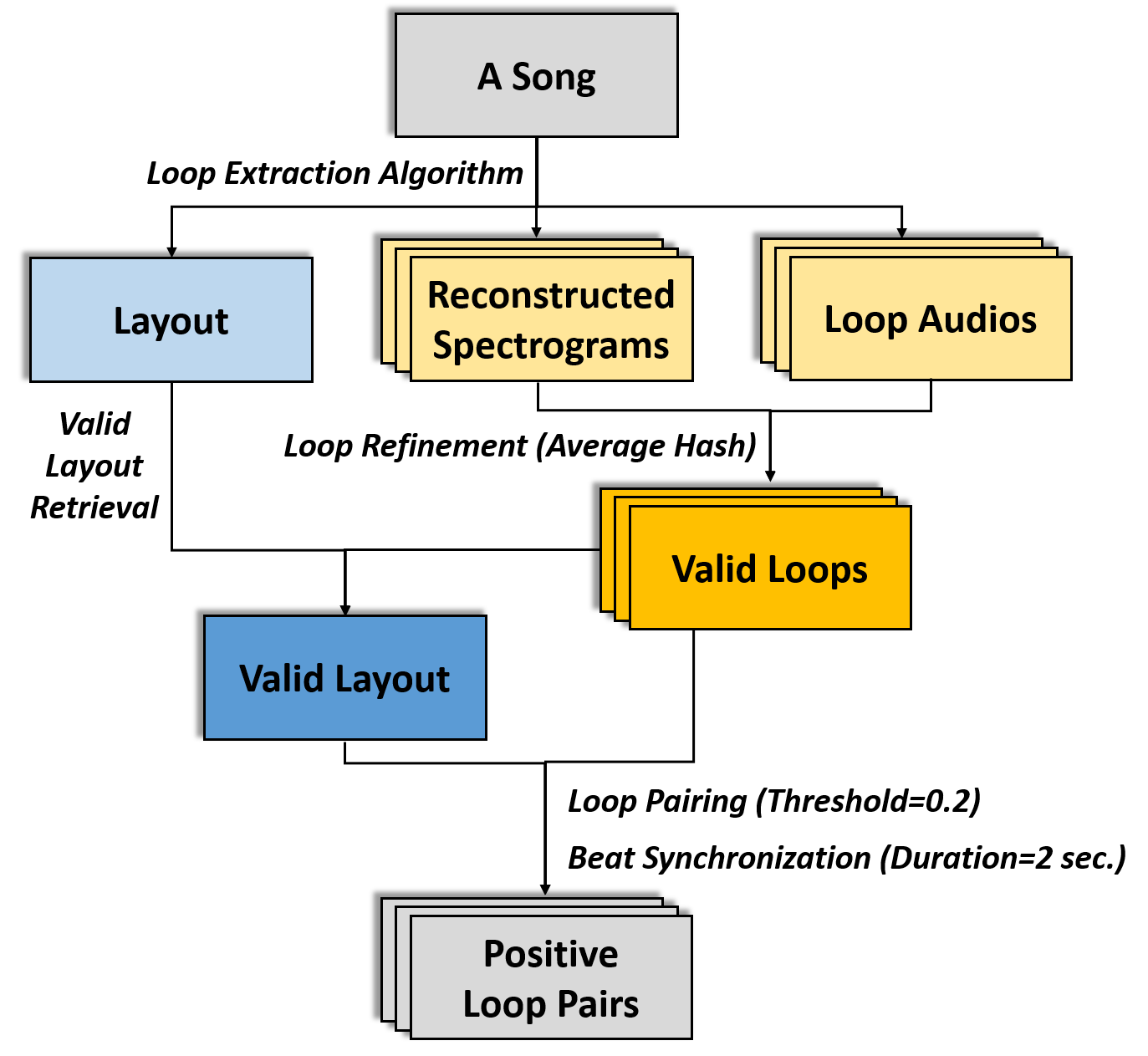}
\caption{The proposed data generation pipeline for creating positive loop pairs.} 
% \st{for use in a supervised machine learning model for estimating loop compatibility}.}
\label{fig:pipline}
\end{figure}

\subsection{Data Generation}
\label{sec:date_generation_method}

Figure \ref{fig:pipline} depicts the overall data generation pipeline.
We firstly use the algorithm proposed by Smith and Goto \cite{18icassp_loop_sepearation} for loop extraction.
The algorithm 
%uses a downbeat tracking algorithm %(like them, we used madmom)
%and a number of loop patterns to search for.
%It then 
uses NTF to model a song as a set of 
sound templates and rhythm templates, \jordan{a set of ``recipes'' for combining these templates to recreate the loop spectrograms,}
and a loop layout that defines when each loop occurs in the song.
%\st{Then, there is an algorithm to decide which combination of sound and rhythm templates to use to create audio loops~\cite{19ismir_unmixer}.}
% (the sound templates, rhythm templates are combined to define each loop spectrogram)
%A process for choosing which spectrograms to invert to obtain the audio is described in~\cite{19ismir_unmixer},
\jordan{In later work, the authors described how to choose the best instance of a loop to extract from a piece, and a second factorization step that can be applied to reduce the redundancy of the loops~\cite{19ismir_unmixer}.}
%\st{Moreover, there is a second factorization step to reduce the redundancy of the loops.}
As a result of this process, we get out of a song  the loop layout, the reconstructed spectrograms, and the audio files of loops, as depicted in the first half of Fig.~\ref{fig:pipline}.

\jordan{Despite this extra step, we still found many redundant, similar-sounding loops from the Hip-Hop dataset. To further deduplicate the loops, we introduce a new loop refinement step.}
The main idea is to consider the reconstructed spectrograms  obtained from the loop extraction algorithms as an image, and apply the average hash algorithm \cite{imagehash} used in the identification or integrity verification of images to detect the duplicate loops. We first construct the hash from each spectrogram extracted from the same song, then count the number of bit positions that are different in every pair of spectrograms. If a pair of spectrograms has fewer than five bit positions that are different, we regard them as duplicates and remove one of them.

% We consider constructed spectrograms as an image and conduct a similarity estimation by the Structural Similarity Index (SSIM Index)\cite{ssim}. If two reconstructed spectrograms are too similar, we view them as duplicate loops and remove one of them. %After that, we can refine most of the duplicate loops and obtain high-quality valid loops. 

After this, we refine the loop layout by two steps. First, we combine all activation values from the duplicate loop in the loop layout into a single activation value. Second, we normalize all the activation values in each bar. This leads a valid loop layout that is ready for loop pairs creation.

Finally, we have to process the real-valued loop layout to obtain pairs of loops that do co-occur. A straightforward approach %to achieve this 
is to threshold the loop layout; we found a threshold of 0.2 to work reasonably well. 
% The outcome is a median of 3 unique loops and 2 unique loop pairs per piece.
Please note that, to make all the loops in our dataset comparable, we time-stretch each to be 2 seconds long.
%\st{The purpose of \textbf{loop pairing} is to have an idea of when loops co-occur in a song.}
%To define positive and negative pairs of loops, we need to know which of them co-occur in each song.
% For the purpose of loop pair creation, we need to clarify which loops have co-occurred in a song. In other words, we need to know how many and which loops are used in each bar of the original song.
%\st{A straightforward approach to achieve this would be to use the estimated loop layout. Yet, the loop layout has to be thresholded first.}
%\st{With a binary layout, we can rapidly create positive loop pairs like this: if $n$ loops co-occur, we get $C^n_2$ pairs.}
%\st{Empirically, we found a threshold of 0.2 to binarize the layout works reasonably well. Please note that, to make all the loops in our dataset comparable, we time-stretch each to be 2 seconds long.}
% To achieve this, we make use of loop layout information which contains loop activation values per bar. Then, we set a  threshold  (0.2) to determine whether the loop is used in a bar. 
%If the activation value is higher than the threshold value, we regard this loop that contributes enough to this bar. 
% Afterward, we create the loop pair by paring all the possibilities in each bar. Finally, we time-stretch all the loops to the same duration (set to be 2 seconds) and synchronize it to consistent beats. % for the further combining purpose.

\section{Proposed Learning Methods}
\label{sec:method}

%\paul{In this section, we describe the five different strategies we have implemented for sampling negative data for model training. For a fair comparison, we set up the ratio of positive  to negative data is 1: 2.
%which are highly correlated with model training. 
%Specifically, we consider the following five strategies and report the results in the next section. Following, 
%Based on a dataset of positive and negative examples, we investigate two supervised neural network-based approaches for estimating compatibility.}

\begin{figure}[t]
\centering
\hspace*{-0.5cm}
\includegraphics[width=7cm,height=6cm,keepaspectratio]{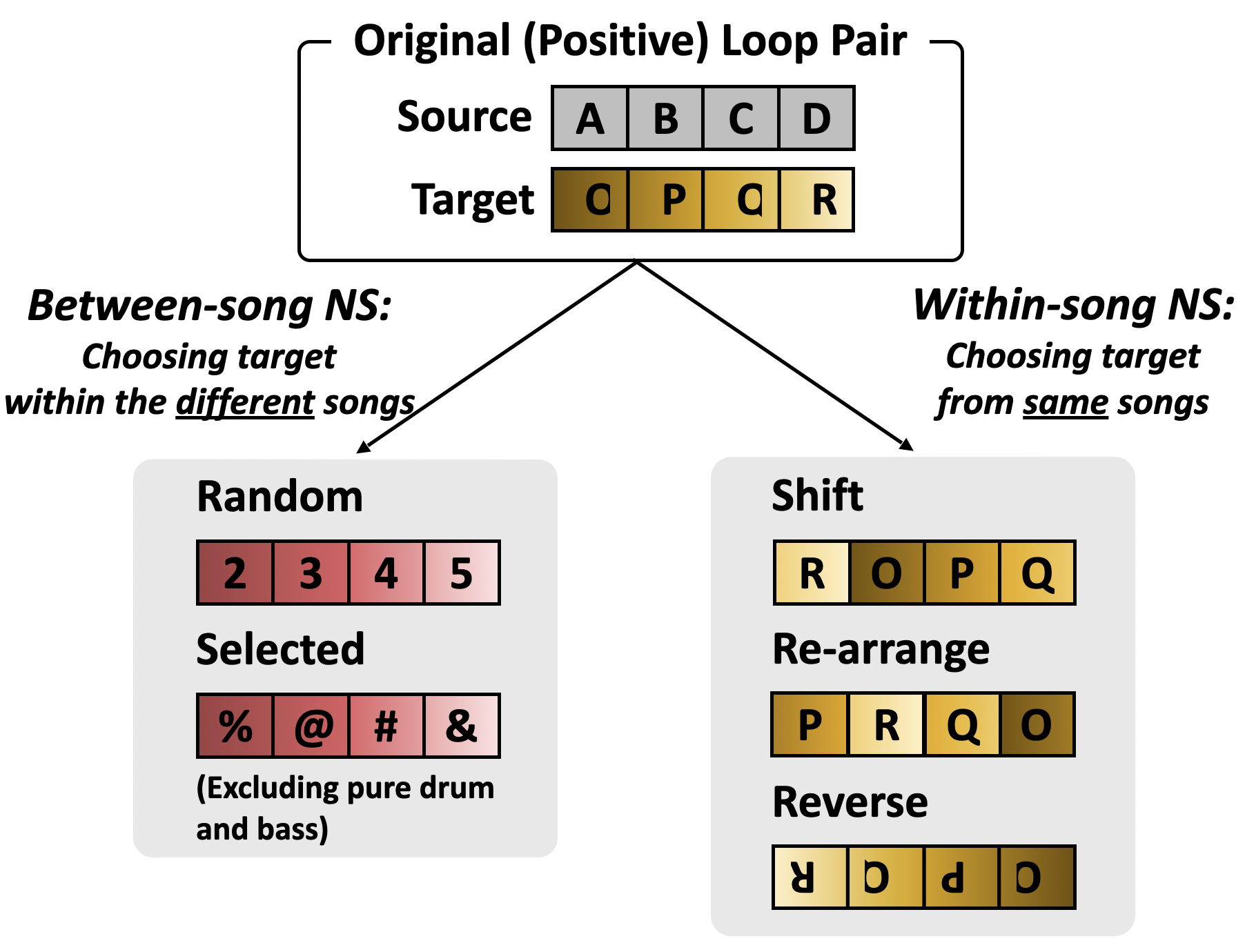}
\caption{Illustration of various  loop-pair `negative sampling' (NS) strategies explored in the paper.}
\label{fig:negative_sampling}
\end{figure}

\subsection{Negative Sampling Strategies}
\label{sec:method_ns}

We get abundant positive data after the data generation pipeline. However,
\jordan{we also need negative examples (pairs of clips known to not match well) to train our mashability predictor.}
% to train a  \paul{supervised} learning model,
% \jordan{binary classifier},
% negative data is also needed.
While there are straightforward 
\jordan{ways to collect such examples},
% solutions,
proper and domain-specific negative sampling has been shown to be important \cite{15iccvfeature,15facenet,17iccvsampling,18speech,18disambiguating}.
%%However, a negative sampling strategy's efficiency differs according to different tasks. 
%For example, distance weighted sampling \cite{17iccvsampling} , Zipf’s Law weighted sampling \cite{18speech}, and artist-based sampling \cite{18disambiguating} have been found useful in image retrieval, speech representation learning, and artist clustering, respectively.
%%propose  which is proven to be efficient in image retrieval and clustering task.  propose that taking Zipf’s Law into account during the negative sampling can help neural network to learn invariant speech representation.  propose to take the side information related to music artist in negative sampling can help address the problem of an unknown artists clustering task. 
%Similarly, for our model, we should provide incompatibility loops as negative data. 
\jordan{We  experiment with the two types}
% We therefore experiment with the two families 
of methods of negative sampling.
See Figure \ref{fig:negative_sampling} for an overview of such methods.
%First, we can make loop pairs as incompatible as possible, which we call \emph{incompatible negative sampling}. 
%Second, contrary to positive data, negative data can be created by pairing loops from different song, which we call \emph{naive negative sampling}. 
%We experiment with 5 negative sampling strategies that contain 2 naive negative samplings and 3 incompatible negative samplings to provide the negative data for our learning system

\subsubsection{Between-Song Negative Sampling} 
\label{subsubsec:naive ns}
A naive approach to negative sampling, dubbed the \textbf{random} method, is to take any two loops from two different songs, and call that a negative pair. But, it is hard to ensure the loops collected in this way clash with one another.
\paul{We therefore also study a slightly more sophisticated method that takes \emph{instrumentation} into account, dubbed the \textbf{selected} method. Specifically, we noted that many loops are pure drum or pure bass loops, and they tend to be compatible with any loops. Therefore, we use \cite{spleeter2019} to detect pure drum or bass loops and avoid selecting them in the process of random negative sampling. This way may help emphasize the harmonic compatibility of the loops. With this strategy, we can experiment whether putting more emphasis on \emph{harmonic compatibility} can indeed improve the result, or other feature's compatibility is still crucial.}

\subsubsection{Within-Song Negative Sampling}
% \subsubsection{incompatible Negative Sampling}
\label{subsubsec:disng}
 
\yh{Negative data can also be created by editing a loop in a positive loop pair.}
\paul{We come up with three methods for making conflicting loop pairs: 
%by modifying the original data:  
`reverse,' `shift,' and `rearrange.' 
Given a positive loop pair, we view one of them as the \emph{source loop}, and the other as the \emph{target loop}.
%Specifically, to create a pair of conflicting loops, we separated a positive example into a \emph{source loop} and a \emph{target loop} in advance, then manipulated the target loop into different degrees of conflicting loops as follows. 
With the \textbf{reverse} method, the target loop is played backward to create the rhythmically and harmonically conflict. The \textbf{shift} method cycle-shifts a target loop by 1 to 3 beats at random to make the source loop and manipulated target loop unbalanced. The \textbf{rearrange} method cuts the target loop into beats and reorders them randomly.} %In this three-way, we can effectively create the loop pairs that sound unusual and conflicting compared to the original data.}
The combination of the source loop and the manipulated target loop is considered as a negative loop pair.
See Figure \ref{fig:negative_sampling} for illustrations.

% \begin{figure*}[t!]
% %\centering
% \begin{subfigure}[t]{0.5\textwidth}
%     \centering
% 	\includegraphics[width=10cm,height=8cm,keepaspectratio]{figs/cnn_model.png}
%     \caption{CNN}
%     \label{fig: CNN}
% \end{subfigure}
% \hfill
% \begin{subfigure}[t]{0.5\textwidth} 
%     \centering
% 	\includegraphics[width=10cm,height=8cm,keepaspectratio]{figs/snn_model.png}
%     \caption{SNN}
%     \label{fig: SNN}
% \end{subfigure}
% \caption{CNN and SNN Model Architectures}
% \label{fig: model}
% \end{figure*}

\begin{figure}[t!]
\centering
\hspace*{-0.3cm}
\includegraphics[width=8cm,height=12cm, keepaspectratio]{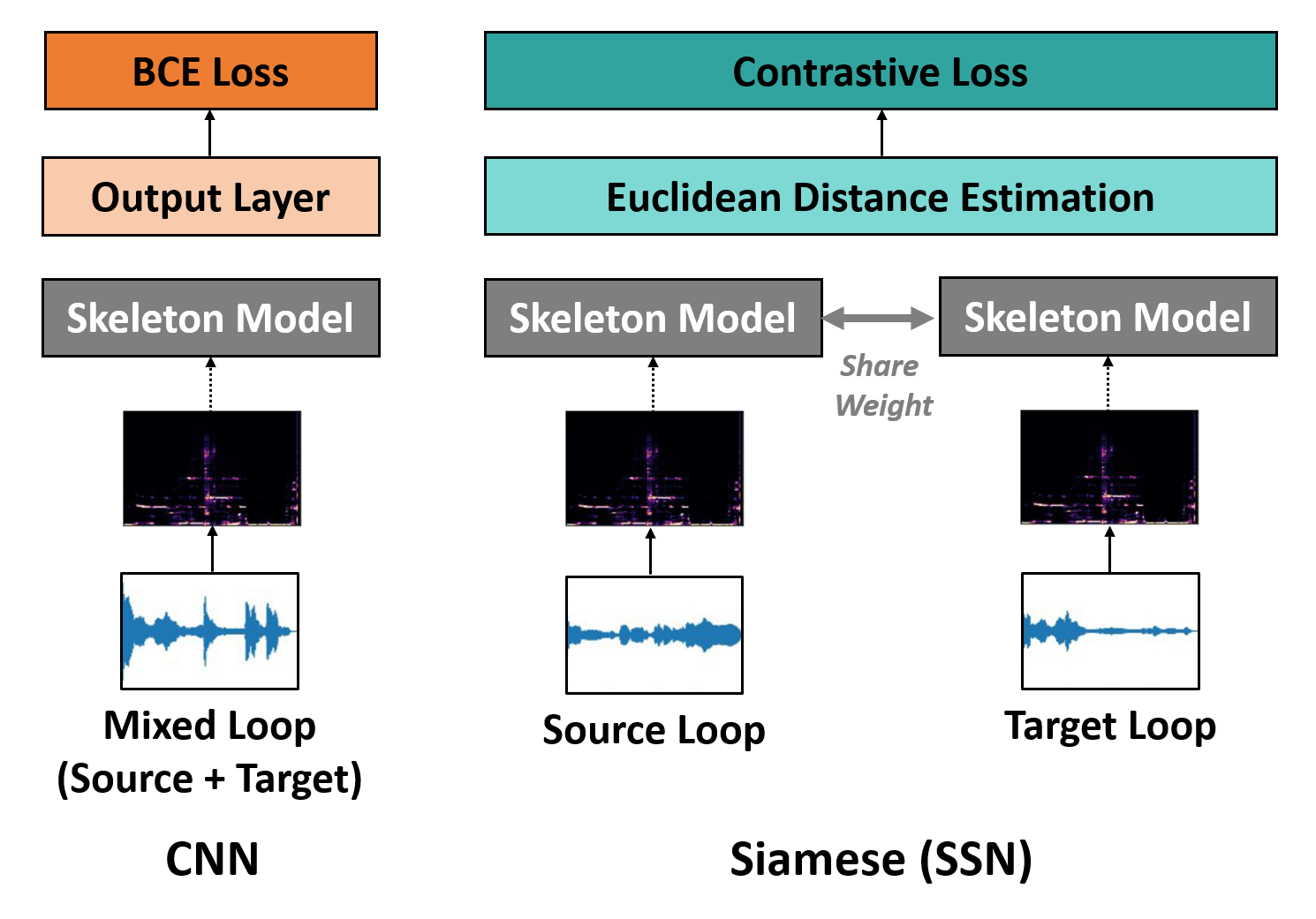}
\caption{Architectures of the CNN and SNN models studied in this paper. The CNN takes a pair of loops (in the time domain) as the input, whereas the SNN takes the two loops as separate inputs and concatenate the intermediate abstractions only in the middle of its network.}
\label{fig: model}
\end{figure}

\subsection{ Model and Training}
\label{sec:method_model}

\jordan{Figure \ref{fig: model} shows the two models (CNN and SNN) proposed to learn the compatibility of two loops. % in different ways. 
To make a fair performance comparison, we train them with the same skeleton in a different way. Therefore, we first introduce the skeleton model and then explain how it is incorporated into the specific model architectures for the training process.}
\subsubsection{\paul{Skeleton model}}
\label{subsubsec:skeleton model}
\paul{A two-layer 2-D convolutional neural network (CNN) and 3-layer fully connected neural network is the skeleton of networks used in this paper. There are 16 filters (with kernel size 3) and 4 filters (also with kernel size 3) in the first and second convolutional layers. The 3-Layer fully-connected neural network is constructed by 256, 128, 16 output features, respectively. Furthermore, batch normalization\cite{batch_norm}, 10\% dropout\cite{droupout}, and PReLU\cite{PReLU} are applied to all convolutional and fully-connected layers. All the models are trained using stochastic gradient descent with batch size 128.}
\paul{The input for the skeleton model is a 2-second loop audio. We compute the spectrograms by sampling the songs at 44k Hz and using a 2,048-point Hamming window and 512-point hop size. We then transform the spectrograms into 128-bin log mel-spectrograms. The resulting input has shape 173 frames by 128 bins.}

\subsubsection{\paul{CNN Model}}
\label{subsubsec:cnn model}
\jordan{A convolutional neural network is well-known for its ability as feature extraction.
% It is common to add a fully-connected layer for the classification purpose.
In our system, in order to learn the compatibility of two loops, we propose to use a CNN as a classifier by combining two loops into a single input, and then train the CNN model to learn to distinguish whether loop combinations would sound harmonious (high compatibility) or incompatible (low compatibility). For the classification purpose, we stack the skeleton model with a fully-connected output layer to get a single value as the output. Then, we compute the binary cross entropy loss (BCELoss) to update the parameters of the whole model.
% After training, the model is expected to have the ability to distinguish harmonious or incompatible loop pairs.
We note that its output is a value between 0 and 1, with values closer to 1 indicating a higher probability that the pair of loops are compatible, and closer to 0 when they are not.
% If the output of the model is close to 1, that means the model believes that loop pairs have a high probability of being harmonious. In contrast, the model tends to regards loop pairs incompatible if its output approximate to 0.
Therefore, we can later use its output to estimate the compatibility of any two loops.
%Later, we will treat the output value from the model as the compatibility value to find suitable loop pairs.
}

% \begin{table*}[t]
% \centering
%     \begin{tabular}{cl|cccc|cc}
%         \toprule
%         \textbf{Model} & \textbf{Negative sampling} & \textbf{Avg rank.}& \textbf{Top-10 acc.}& \textbf{Top-30 acc.}& \textbf{Top-50 acc.}&\textbf{* Acc.}&\textbf{* F1.} \\ 
%         \midrule
%         \multirow{2}{*} 
%         &Random&{42.96}&{0.13}&{0.35}&{0.59}&{0.60}&{0.59} \\ 
%         &Selected&{43.13}&{0.13}&{0.29}&{0.62}&{0.59}&{0.59}\\ 
%         {CNN}&Reverse&{41.19}&{0.19}&{0.42}&{0.62}&{0.63}&{0.62}\\
%         &Shift&{48.97}&{0.11}&{0.34}&{0.54}&{0.57}&{0.56}\\
%         &Rearrange&{47.72}&{0.10}&{0.31}&{0.57}&{0.57}&{0.57}\\\midrule
%         &Random&{34.18}&{0.27}&{0.52}&{0.74}&{0.51}&{0.47}\\
%         &Selected&{42.75}&{0.18}&{0.39}&{0.59}&{0.52}&{0.47}\\
%         SNN&Reverse&{42.72}&{0.16}&{0.37}&{0.62}&{0.53}&{0.48}\\
%         &Shift&{42.95}&{0.16}&{0.41}&{0.65}&{0.53}&{0.52}\\ 
%         &Rearrange&{44.20}&{0.16}&{0.40}&{0.60}&{0.53}&{0.53}\\
%         \bottomrule
%     \end{tabular}
%     \caption{Evaluation results of the classification method (Average rank and Top accuracy) and ranking method (Accuracy and F1-Score).}
%     \label{tab:results}
% \end{table*}

\subsubsection{Siamese Model}
\label{subsubsec:siame cnn model}

A Siamese neural network (SNN) \cite{94siamese} consists of twin networks that share weights and configurations. 
SNN has been shown effective in image retrieval  \cite{15siamese} and various MIR tasks alike \cite{19ismir_voice, 17ismir_artist, 18aaai_playing_puzzle,Ubai17tist,manocha18icassp,yu2017hit}. Aiming to test its applicability to loop compatiblity estimation, we train an SNN using the labeled data we created through the data generation pipeline as follows.
%We train an SNN model with sufficient positive and negative data. 
The outcome of an SNN is a mapping function from the input features to the output vectors in an embedding space. 
During the training process, our SNN first transforms the input Mel-spectrograms into a vector by a skeleton model and then optimize in contrastive loss\cite{06cvpr_contrastive_loss} directly. After training, we have a mapping function that can map any loop to the embedding space. To select the compatible loops, we compute the Euclidean distance between two loops in that embedding space. If the distance for two loops is close, then we assume they may be compatible, and vice versa. %In contrast, If the distance is far, two loops may conflict with each other.

% \begin{table*}[t]
% \centering 
%     \begin{tabular}{cl|cccc|cc}
%         \toprule
%         \textbf{Model} & \textbf{Negative sampling} & \textbf{Avg rank.}& \textbf{Top-10 acc.}& \textbf{Top-30 acc.}& \textbf{Top-50 acc.}&\textbf{* acc.}&\textbf{* F1.} \\ 
%         \midrule
%         \multirow{2}{*} 
%         &Random&{42.96}&{0.13}&{0.35}&{0.59}&{0.60}&{0.59} \\ 
%         &Selected&{43.13}&{0.13}&{0.29}&{0.62}&{0.59}&{0.59}\\ 
%         {CNN}&Reverse&{41.19}&{0.19}&{0.42}&{0.62}&{0.63}&{0.62}\\
%         &Shift&{48.97}&{0.11}&{0.34}&{0.54}&{0.57}&{0.56}\\
%         &Rearrange&{47.72}&{0.10}&{0.31}&{0.57}&{0.57}&{0.57}\\\midrule
%         &Random&{34.18}&{0.27}&{0.52}&{0.74}&{0.51}&{0.47}\\
%         &Selected&{42.75}&{0.18}&{0.39}&{0.59}&{0.52}&{0.47}\\
%         SNN&Reverse&{42.72}&{0.16}&{0.37}&{0.62}&{0.53}&{0.48}\\
%         &Shift&{42.95}&{0.16}&{0.41}&{0.65}&{0.53}&{0.52}\\ 
%         &Rearrange&{44.20}&{0.16}&{0.40}&{0.60}&{0.53}&{0.53}\\
%         \bottomrule
%     \end{tabular}
%     \caption{Results of objective evaluation. }
%     \label{tab:results}
% \end{table*}

\begin{table*}
\centering
    \scalebox{.9}{
    \begin{tabular}{ll|cc|cccc}
        \toprule
        \multirow{2}{*}{\textbf{Model}} & \textbf{Negative} &  \multicolumn{2}{c|}{\textbf{Classification-based metric}} & \multicolumn{4}{c}{\textbf{Ranking-based metric}}  \\ 
         & \textbf{sampling} &\textbf{Accuracy}&\textbf{F1 score} & \textbf{Avg. rank}& \textbf{Top 10}& \textbf{Top 30}& \textbf{Top 50}\\ 
        %& \textbf{}& \textbf{10}& \textbf{30}& \textbf{50}&\textbf{acc.}&\textbf{F1.} \\ 
        \midrule
            & Random&{0.60}&{0.59} &{43.0}&{0.13}&{0.35}&{0.59}\\ 
            & Selected&{0.59}&{0.59}&{43.1}&{0.13}&{0.29}&{0.62}\\ 
            \cline{2-8}
        CNN & Reverse&{\textbf{0.63}}&{\textbf{0.62}}&{41.2}&{0.19}&{0.42}&{0.62}\\
            & Shift&{0.57}&{0.56}&{49.0}&{0.11}&{0.34}&{0.54}\\
            & Rearrange &{0.57}&{0.57}&{47.7}&{0.10}&{0.31}&{0.57}\\\midrule
            & Random&{0.51}&{0.47}&{\textbf{34.2}}&{\textbf{0.27}}&{\textbf{0.52}}&{\textbf{0.74}}\\
            & Selected&{0.52}&{0.47}&{42.8}&{0.18}&{0.39}&{0.59}\\
            \cline{2-8}
        Siamese NN & Reverse&{0.53}&{0.48}&{42.7}&{0.16}&{0.37}&{0.62}\\
            & Shift&{0.53}&{0.52}&{43.0}&{0.16}&{0.41}&{0.65}\\ 
            & Rearrange &{0.53}&{0.53}&{44.2}&{0.16}&{0.40}&{0.60}\\
        \bottomrule
    \end{tabular}
    }
    \caption{Objective evaluation result of different combinations of models (CNN or SNN; see Section \ref{sec:method_model}) and negative sampling strategies (see Section \ref{sec:method_ns}). We highlight the best result for each metric in bold. }
    \label{tab:results}
\end{table*}

\section{ Experiments and Evaluation}
\label{sec:exp and eval}

\subsection{ Objective Evaluation}
\label{obj}

In the objective evaluation, we aim to evaluate the performance of different combinations of model architectures (i.e., CNN and SNN) and negative sampling methods. 
In doing so, we consider two types of objective metrics. 

The first type of evaluation entails a \emph{classification} task. It assesses a model's \jordan{ability to distinguish}
% capability in distinguishing
compatible loops from incompatible ones. 
%pairs from the incompatible ones. 
%based on the training purpose of the CNN model as a classifier. 
To create a test set for this evaluation, we used the positive data from the validation data and collected the negative data by using all the negative sampling methods equally, in order not to favor any of them in this evaluation. 
To make the test set balanced, we set the ratio of negative to positive data to 1:1. %During the first evaluation process, the models were asked to predict how possible the data are compatible, which 1 represents harmonic and 0 represents incompatible. The classification accuracy of both models were calculated and compared.

The second type of evaluation, on the other hand, involves a \emph{ranking} task.
Given a query loop and a certain number of candidate loops, a model has to rank the candidate loops in descending order of
% the estimated
compatibility with the query. We create\jordan{d} the set of candidate loops for a query loop such that we kn\jordan{e}w one and only one of the candidate loops form\jordan{ed} a positive loop pair with the query loop. We \jordan{could}
% can
then evaluate the performance of a model by checking the position of this ``target loop'' in the ranked list.
\jordan{The closer the rank is to 1, the better.}
% The rank is of course the higher the better.
This evaluation task aligns well with the real-world
\jordan{use case}
% usage
of the proposed model: to find loops compatible with a query \jordan{loop}
% lab in hand
from a pool of loop libraries. However, we note that it may not be always possible to rank the target loop high in this evaluation, because the other loops in the candidate pool may also be compatible with the query. 

%\paul{To evaluate the ability to learn loop compatibility from original loop pairs based on the characteristic of SNN, we recorded 2 models' ranking results of hand-made loop pairs to test whether they are able to rank the existing loop pairs as high as possible. In this phase, Suffering from the absence of compatibility ground-truth value, we regarded the existing loops as a ground truth data. We used the test set containing 100 unique positive pairs in the whole test loops collection. Given a source loop, the models then sorted all the target loops by their compatibility value with the source loop which is descending and ascending in CNN and SNN model respectively. Then, 
In our experiment, we set the number of candidate loops to 100.
\jordan{We computed four metrics for the ranking task: top-10 accuracy, top-30 accuracy, and top-50 accuracy---which evaluate whether the target loop was placed in the top-10, top-30, and top-50 of the ranked list, respectively---as well as the average rank.
We report the average of these values for the 100 different query loops.}

\begin{figure*}[t!]
\centering
\hspace*{-0.5cm}
    \includegraphics[width=18cm,height=15cm,keepaspectratio]{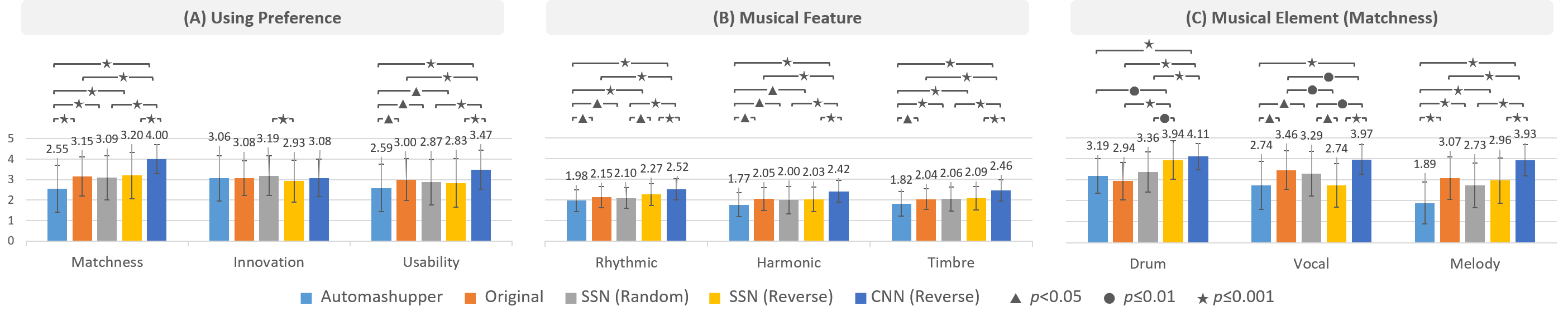}
\caption{The subjective evaluation results of comparing the preference, musical feature and musical element among 5 methods. A method is considered to show low abilities to manipulate loops if its MOS is under 3 in (A) `using preference'-related and (C) `musical elements'-related metrics, and under 2 in (B) `musical feature'-related metrics. }
\label{fig: us}
\end{figure*}

\textbf{Classification Result}---Table \ref{tab:results} shows that, regardless of the negative sampling method, the CNN models outperform the SNN models for the classification-based metrics.
\jordan{While the CNN learns to rate the compatibility of the loop combinations directly, the SNN learns to place the loops in a space, with nearby loops being more compatible. The training data were the same, so the difference in performance
suggests that the space learned was not effective for the classification task.}
% may tell us that the space learned by the SNN is not an effective projection.}
% This result may make sense, as the SNN model is not trained for the classification purpose.
% In other words, the SNN may not be as good at classification globally, but it is possible that the SNN does a good job in finding suitable loops locally. This is reflected in the result of the ranking-based evaluation.

%and the reverse negative sampling achieve similar performance with random, selected negative sampling. 

\textbf{Ranking Result}---Table \ref{tab:results} also shows CNN and SNN models seem to perform comparably with respect to the ranking-based metrics. Yet, the best scores in the four ranking-based metrics are all obtained by SNN with `random' negative sampling. 
% Interestingly, 
When \jordan{an} SNN is used, there appears to be a large performance gap between `random' and all the other negative sampling methods. 
This suggests that focusing on harmonic compatibility alone (e.g., as done by using the `selected' negative sampling method) \jordan{is not optimal;}
% does not perform well. The
compatibility in other dimensions of music is also important. 
On the other hand, when \jordan{a} CNN is used, `reverse' appears to perform the best among the five negative sampling methods. 
%Furthermore, all incompatible negative sampling does also not improve the ability of SNN. Surprisingly, comparing with another negative sampling, reverse negative sampling achieves the best performance in CNN which is even better than the random and selected negative sampling in CNN. It is consistent with the classification results. Therefore, we argue that the reverse negative sampling guides CNN to train in a pleasing way.

\subsection{ Subjective Listening Test}
\label{sub}

%\subsubsection{Setup}

We note that even if a model obtains the highest classification or ranking result, we still cannot guarantee that the model also works well in real-world applications. Accordingly, we deploy\jordan{ed} a user study by releasing an online questionnaire to get user feedback. As the users' time is precious, it is not possible to test the result of all combinations of the models and negative sampling methods. Therefore, we consider\jordan{ed} the result of the objective evaluation and pick\jordan{ed} five methods for subjective evaluation:
\begin{itemize}
	\setlength\itemsep{-.4em}
    \item `CNN + reverse', `SNN + random,' and `SNN + reverse,' three instances of the proposed
    % model-based 
    methods
    that performed well in the objective evaluation; 
    \item `AutoMashUpper' \cite{14aslp_atp}, as it represents the state-of-the-art for the task.  \yh{Our implementation is based on the open source code from 
    % [Online]
    \url{https://github.com/migperfer/AutoMashupper}. As the rhythmic compatibility part of \cite{14aslp_atp} is missing in this repo, we implement it ourselves following \cite{14aslp_atp}.}
    \item \yh{`Original,' which stands for \jordan{the} real loop combinations observed in FMA and extracted by the procedures described in Section \ref{sec:date_generation_method}. Specifically, an `original' loop combination is one of the 100 positive loop pairs from the `test set' listed in Table \ref{tab:numbers}.}
\end{itemize}
%Furthermore, to assess the influence of the different negative sampling methods, we also include in our subjective evaluation the  which also performs well in the ranking-based evaluation. Lastly, to compare our system to the state of the art, we also present the suggested loop combinations of AutoMashUpper \cite{14aslp_atp} to the users. 

\yh{ %Specifically, 
The loop combinations, or \emph{test clips}, presented to users for evaluation are created as follows. As in Section \ref{subsubsec:disng}, for each positive loop pair, we view one
% of them
as the source loop, and the other as the target loop. Accordingly, we have 100 source loops and 100 target loops in the test set.
% To form a test clip, the `original' method use the source loop and target loop from the same loop pair, whereas the other four methods search among all the 100 target loops to find the one that is considered the best match (by that method) for a given source loop.
\jordan{A test clip is a combination of two loops. For the `original' method, we pair the source loop with its true target. For the other methods, the 99 target loops (excluding the true target) are ranked, and the one predicted to match the source best is combined with it to form a test clip.}
The 5 resulting test clips for each source loop were considered as a \emph{group}. This way, we have 100 groups in total. Among them, we picked 6 groups for evaluation: they have 2 vocal loops, 2 drum loops, and 2 loops of instrumental melody as the source loop, respectively. }

% \jordan{To form a test clip from a given source loop, each of the four methods is used to rank the 100 target loops (excluding the true matching to find the one that is estimated to match the source best.}

%In order to evaluate the methods' abilities to pair different musical elements, we follow the following procedure to prepare the test clips. 

\paul{We designed a subjective listening test that could be administered to users through an anonymous online questionnaire, and advertised it on social media websites related to EDM and Hip-Hop.
% The 5 systems we compared are: AutoMashUpper, 
% that enjoy listening to or composing loop-based music. 
A participant was randomly directed to a questionnaire containing one group of test clips, and was then asked to listen to the 5 test clips and rate each clip, on a 5-point Likert scale, in terms of: i) whether they sounded \emph{matched}, ii) whether they were an \emph{innovative} combination, iii) and whether they were \emph{usable} in future compositions (see Figure \ref{fig: us}A). They were also asked
to indicate whether the loops matched according to 3 musical features: rhythm, harmony, and timbre (see Figure \ref{fig: us}B).
% to comment on which parts of the audio matched or did not match in 3 musical features.
}
\jordan{Finally, to see how consistent the models are at recommending loops given different query types, Figure \ref{fig: us}C breaks down the results for Matchness (Figure \ref{fig: us}A(i)) according to source loop type: drum, vocal or melody loop.}
% \yh{Finally, we also break down the result in \emph{Matchness} according to whether the source loop is a drum, vocal, or melody loop (see Figure \ref{fig: us}C).}

% \begin{figure}[t!]
% \centering
% \hspace*{-0.5cm}
% \includegraphics[width=9cm,height=12cm,keepaspectratio]{figs/us_result_01.png}
% \caption{The result of subjective evaluation comparing the pairing abilities of match, innovation and usability among 5 methods. (`$\bigtriangleup$' represents $p < 0.05$; `$\bullet$' represents $p \leq 0.01$; `$\star$' represents $p \leq 0.001$)}
% \label{fig: us_01}
% \end{figure}

\subsubsection{Subjective Evaluation Result}

Data from 116 \yh{Taiwanese} \jordan{participants} were collected, and all of them were included for analysis. The participants self-reported the gender they identified with (36 female, 80 male) and age (19--33). 50\% said they listened to loop-based music more than 5 days a week, 54\% had classical musical training, and 42\% had experience composing loop-based music.
Overall, the responses indicated an acceptable level of reliability (Cronbach’s $\alpha=0.778$).

\paul{Figure \ref{fig: us} indicates the mean opinion score (MOS).
%and the corresponding $p$-value of each method in user preferences, musical features and elements. 
A Wilcoxon signed-rank test
% \cite{wilcoxon}
was conducted to statistically evaluate the comparative ability of the 5 methods.}

Overall, we observe that \yh{AutoMashUpper performs \jordan{least well}
in almost all the evaluation metrics. Outside of the \emph{Innovation} (Figure \ref{fig: us}A) and \emph{Drum} metrics (Figure \ref{fig: us}C), AutoMashUpper is significantly worse than almost all the other methods. This suggests that, in loop selection, considering only loop similarity \jordan{and spectral balance} (as assumed by AutoMashUpper) is not enough---perceptually compatible loops are not necessarily similar in content.}

On the other hand, we note that \yh{`CNN with reverse negative sampling' performs the best in almost all of the evaluation metrics.
\jordan{To our surprise, the loop combinations picked by the CNN are preferred even to the original loop combinations extracted from FMA songs.}
The performance difference between `CNN + reverse' and `Original' is significant ($p$-value $<$ 0.001) in terms of \jordan{several metrics, including} \emph{Matchness} and \emph{Usability}.
In contrast, there is no significant performance difference between `Original' and the two SNN models.}

One finding was surprising: \yh{The MOS \jordan{obtained by} `CNN + reverse' for \emph{Matchness} is 4.0,
% which is quite high on a 5-point scale, and
higher even than the MOS of 3.15 obtained by `Original.' I.e., the loop combinations proposed by the system were found to match better than the original combinations.}
\jordan{We are not sure how to account for this success. One might conjecture that the quality of the `Original' pair suffers from the imperfect loop extraction algorithm, but since all the loops were extracted the same way, they should be on equal footing.
Alternatively, the novelty of the non-original mash-ups could be more interesting to the listeners than the originals; however, there was no clear difference among the systems in terms of `Innovation.'}
\yh{We can only conjecture that `CNN + reverse' found better loop combinations than human-made ones because the model-based method could examine all the possible loop combinations, while humans cannot.}

We found that 
\yh{SSN methods, despite their performing better in ranking-based objective metrics than the CNN method, did not create perceptually-better loop combinations. This suggests a discrepancy between the  objective and subjective metrics.}
\yh{And lastly, while it seems fair to say that CNNs outperform SNNs in the subjective evaluation, it is hard to say whether `reverse' is the most effective negative sampling, because we were not able to evaluate all the possible methods here. This is left as a future work.}

\section{Conclusion and Future Work}
\label{sec:conclusion}

In this paper, we have proposed a data generation pipeline and several negative sampling strategies to address the need of ground-truth labels for building a machine learning model for estimating the compatibility of loops. We have also implemented and evaluated two different network architectures \jordan{(CNN and SNN) to build such models}.
% , namely CNN and Siamese NN, to build such models. 
Our objective evaluation shows that \jordan{a} CNN does well in classifying incompatible loop pairs and \jordan{an} SNN is good at imitating how producers combined loops, and our subjective evaluation suggests the loop combinations created by \jordan{a} CNN are favored over those created by \jordan{an} SNN and even\jordan{, in some aspects,} real data from FMA. Both CNN\jordan{s} and SNN\jordan{s} outperform the rule-based system AutoMashUpper.
%To consider both objective and subjective evaluation results, we found that SNN architecture can imitate the way how the previous producer combines the two loops. However, the effectiveness might be limit by the positive loop pairs which provide by the training data. On the other hand, we discover that CNN architecture can not only understand why these two loops used to combined before but not limit to the training data. CNN provides the new loop combination suggestion which may be better than the original one which is pairing by the real producer.}
%In this paper, td  several methods for ..

\jordan{We have two plans in place for future work.}
% We have the following two more plans in place for future work.
%considering the inconvenience of extracting loops from a large-scale database and
%This work would be extended in the future in the following aspects.   Furthermore, we verify that reverse negative sampling work well on both models based on the subjective evaluation results.
%Powered by the open-source loop extraction algorithm, our approach is available for every researcher by following the instructions in this paper. 
First, as we see some inconsistency between the results of objective and subjective evaluations, we \jordan{plan} to investigate other objective metrics for performance evaluation. 
%Therefore, to meet user's preferences in real-life scenarios, building a dataset with the compatibility annotation through artificial annotation or using data tracking from interaction interfaces is needed.
Second, we plan to exploit the loop layouts estimated by the loop extraction algorithm \cite{18icassp_loop_sepearation} to study further the relationship between loops and their arrangement, which may
\jordan{aid in the}
% contribute to
automatic creation of loop-based music. %that has long-term structure.

\section{Acknowledgement}
% Hidden for blind review.
This research was in part funded by a grant from the Ministry of Science and Technology, Taiwan (MOST107-2221-E-001-013-MY2).

%\vspace{-1em}
\bibliography{ref}

\end{document}